\documentclass[proof]{pasj01}

\Received{$\langle$reception date$\rangle$}
\Accepted{$\langle$acception date$\rangle$}
\Published{$\langle$publication date$\rangle$}

\usepackage{ulem}

\begin{document}

\title{Energy estimation of high energy particles associated with the SS433/W50 system through radio observation at 1.4 GHz}
\author{Haruka \textsc{Sakemi}$^{1,*}$, \and Rikuto \textsc{Omae}$^{1}$, \and Takumi \textsc{Ohmura}$^{1}$, \and \\and Mami \textsc{Machida}$^{2}$}%
\altaffiltext{1}{Graduate School of Science, Kyushu University, 744 Motooka Nishi-ku, Fukuoka, Fukuoka 819-0395, Japan}
\altaffiltext{2}{National Astronomical Observatory of Japan, 2-21-1 Osawa, Mitaka, Tokyo 181-8588, Japan}

\email{sakemi@phys.kyushu-u.ac.jp}

\KeyWords{{ISM: individual (W50)}$_1$ --- {ISM: jets and outflows}$_2$ --- {ISM: atoms}$_3$ --- {stars: individual (SS433)}$_4$}

\maketitle

\begin{abstract}
The radio nebula W50 is a unique object interacting with the jets of the microquasar SS433. The SS433/W50 system is a good target for investigating the energy of cosmic-ray particles accelerated by galactic jets. We report observations of radio nebula W50 conducted with the NSF's Karl G. Jansky Very Large Array (VLA) in the L band (1.0 -- 2.0 GHz). We investigate the secular change of W50 on the basis of the observations in 1984, 1996, and 2017, and find that most of its structures were stable for 33 years. We revise the upper limit velocity of the eastern terminal filament by half to 0.023$c$ assuming a distance of 5.5 kpc. We also analyze the observational data of the Arecibo Observatory 305-m telescope and identify the HI cavity around W50 in the velocity range 33.77 km s$^{-1}$ -- 55.85 km s$^{-1}$. From this result, we estimate the maximum energy of the cosmic-ray protons accelerated by the jet terminal region to be above 10$^{15.5}$ eV. We also use the luminosity of the gamma-rays in the range 0.5 -- 10 GeV to estimate the total energy of accelerated protons below 5.2 $\times$ 10$^{48}$ erg.

\end{abstract}

\section{Introduction}
\label{sec:intro}

The microquasar SS433 is one of the most famous jet sources in our galaxy. Its outstanding feature is ejection of two-sided precessing jets. The jet velocity is 0.26 $c$ and the precession period is 162 days \citep{abell1979,margon1984,margon1989,davydov2008}. The half-opening angle of the precessing jets is 20$^{\circ}$ \citep{margon1984}. The helical shape of the jets is believed to be broken down within the subparsec scale, and the hollow jets proceed for dozens of parsecs \citep{brinkmann2007,monceau-baroux2015,panferov2017}.

SS433 is located at the geometric center of the large radio nebula W50. The nebula consists of one spherical shell and two elongated structures called ``wings'' \citep{geldzahler1980,downes1981a,downes1981b,downes1986,dubner1998}. Because the jet axis is well aligned with the direction of the wings, the jet and nebula are considered to have some physical association. \citet{elston1987} first observed W50 with the historical Very Large Array (VLA) with a high resolution of 60$^{\prime\prime}$, and suggested that the wings were formed by the ram pressure of the SS433 jet to the W50 shell. \citet{dubner1998} carried out high-sensitivity mosaic observation of W50 with the VLA at 1.4 GHz and succeeded in resolving the complex structures with a high dynamic range. In recent years, wide-band and high-sensitivity full-stokes observation has enabled detailed polarization analysis \citep{farnes2017}. \citet{farnes2017} identified a large-scale rotation measure (RM) gradient across the spherical shell in the east-west direction and across the eastern edge of the wing in the north-south direction using Australia Telescope Compact Array (ATCA) data at 1.1 -- 3.1 GHz. \citet{sakemi2018a} focused on the eastern edge of the wing and identified the helical magnetic fields coiling around the eastern wing. The latest radio continuum observation was reported by \citet{broderick2018}. They observed W50 with the Low-Frequency Array (LOFAR) over the frequency range 115 -- 189 MHz. The morphology of W50 observed at low frequency showed agreement with that of high-frequency observation. They found that only the flux of the eastern wing was lower than the value predicted by the power-law spectrum, probably because of the foreground free-free absorption.

One of the outstanding issues of the SS433/W50 system is the origin of the spherical shell of W50. Two major models are (i) a wind bubble from a binary companion of SS433, a progenitor star, or an accretion disk with a mass accretion rate exceeding the Eddington limit \cite{konigle1983,asahina2014}; and (ii) a supernova remnant (SNR) originating from SS433 \cite{downes1986}. A combination of these models is also possible \cite{ciotti1989,tenorio-tagle1991,rozyczka1993}. However, both scenarios have the same difficulty; although it takes a long time for the shell radius to increase by about 50 pc, the shell is still bright. Furthermore, the mechanism of formation of the wings by the jets has not been determined yet. \citet{goodall2011a} estimated the upper velocity of the eastern terminal region of the jet as 0.0405 $c$ (12,142 km s$^{-1}$) using radio observation datasets for the 12 years between 1984 and 1996. They concluded that a continuous, non-multi-episodic jet could not be decelerated to a velocity below this upper limit because the surrounding interstellar medium (ISM) has a low density of 0.1 -- 1 particles cm$^{-3}$, and intermittent jets are required for the morphology of W50 including the wings \citep{goodall2011b}. However, \citet{panferov2017} suggested that a continuous jet could be decelerated by ambient matter considering viscosity, which means that a continuous jet could construct the wing structures.

Recently, another formation scenario for W50 has been proposed: only jets from SS433 can produce the whole morphology of W50 \citep{ohmura2021}. Some active galactic nuclei (AGNs) show a shell-like structure with wings \citep{roberts2015}, and some numerical simulations have succeeded in reproducing this unique morphology \citep{gaibler2009,hodges2011,rossi2017}. A similar model can be applied to galactic X-ray binary jets considering this scaling law. \citet{ohmura2021} suggested that the morphology of W50 could be reproduced by low-density and continuous jets on the basis of MHD simulation. Their results imply that we should introduce a new hypothesis for the origin of the SS433/W50 system.

In recent decades, microquasar jets have been considered as candidates for accelerators of high-energy cosmic-ray particles above 10$^{15}$ eV \cite{romero2008,pepe2015,cooper2020,sudoh2020}. The SS433/W50 system is an appropriate target for observing cosmic-ray particle acceleration by galactic jets. X-ray observations have identified emission from the SS433 jet \citep{brinkmann1996,safi-harb1997,safi-harb1999,brinkmann2007}. The X-ray hotspot is located near the border between the spherical shell and the eastern wing, and it has both thermal and non-thermal components. TeV gamma-ray emission has been identified at the same position by \citet{abeysekara2018}, who suggested that the origin of the emission was the interaction between cosmic microwave background photons and accelerated electrons through inverse-Compton scattering; this is called the leptonic scenario. \citet{sudoh2020} also support the leptonic scenario theoretically. However, \citet{kimura2020} suggested the hadronic scenario, whereby interaction between cosmic-ray protons and interstellar protons can also explain the TeV gamma-ray emission. 

GeV gamma-ray emission has also been detected by analyzing {\it Fermi}-LAT data over 10 years, and there is still debate on whether it has the same origin as TeV gamma-ray emission. \citet{sun2019} suggested that the position of the GeV gamma-ray emission is different from that of the TeV gamma-rays and is located in the northern area of the spherical shell of W50. They consider the origin of this emission to be the interaction between the spherical shell and the surrounding ISM (hadronic scenario). \citet{li2020} also analyzed the {\it Fermi}-LAT data while considering the influence of the nearby pulsar PSR J1907+0602, and they detected a GeV gamma-ray excess in the northeast area of the spherical shell. They found a periodic variation of this emission with the precession period of the SS433 jet. However, \citet{fang2020} suggested that the position of the GeV gamma-ray emission is the same as that of the TeV gamma-rays, and that these emissions are induced by the SS433 jet. Identifying the origin of the GeV gamma-ray emission is necessary for understanding the cosmic-ray particle acceleration by the SS433/W50 system.

The SS433/W50 system also has another candidate for inducing cosmic-ray particle acceleration at the eastern edge of W50. This candidate is a filamentary structure that is bright in the radio and X-ray bands and perpendicular to the SS433 jet axis \citep{elston1987,brinkmann2007}. In this paper, we call this filamentary structure the ``eastern terminal filament (ETF)''. Because the SS433 jet is believed to reach the position of the ETF, a terminal shock should form in this region. This makes it possible for cosmic-ray particles to be accelerated by the terminal shock. Although no gamma-ray emission has ever been observed here, a future high-sensitivity telescope like the Cherenkov Telescope Array (CTA) may detect it. Verifying cosmic-ray particle acceleration in this region is essential for comprehensively understanding acceleration by galactic jets.

It is difficult to determine the distance of the SS433/W50 system from us. The distance has been estimated at 5.5 kpc on the basis of a kinematic model derived from optical line observation and radio images of the precessing jet \citep{eikenberry2001,blundell2004,blundell2018}. This is different from the distance of $\sim$ 4.6 kpc estimated from parallax observation of the binary companion star with Gaia Data Release 2 \citep{gaia2018}. Recently, the distance has been newly estimated as 8.5 kpc using Gaia Early Data Release 3 (EDR3) \citep{gaia2020}. Therefore, there is a large uncertainty in the distance of SS433 and W50. Because EDR3 still has some calibration issues, we do not consider its value for the distance in this paper \citep{maizapell2021}.

The other method for estimating the distance is to identify the ISM interacting with W50. If we assume the velocity field of our galaxy, we can investigate distances of ISMs on the basis of their line-of-sight velocities. Many studies in recent decades have suggested a relation between W50 and the ISM. \citet{dubner1998} found the surrounding HI gas at the systematic radial velocity $v_{\rm LSR}$ = 42 km s$^{-1}$. The HI gas in their observation seemed to have a cavity-like structure surrounding W50; however, a spatial coincidence with W50 was hard to ascertain owing to a low angular resolution of 21$^{\prime}$. Furthermore, while the velocity center of the HI cavity was revealed, the velocity range of the cavity was still unknown. \citet{lockman2007} detected the HI absorption line of SS433 at a velocity of 75 km s$^{-1}$ and identified the HI emission at the same velocity, which shows good spatial coincidence with the western wing. Molecular clouds were also detected around W50. For example, \citet{yamamoto2008} found molecular clouds whose velocity range was 42 -- 56 km s$^{-1}$. These clouds are well aligned close to the SS433 jet axis. \citet{liu2020} observed a few clouds detected by \citet{yamamoto2008} and suggested that one of them was within the western wing of W50. \citet{su2018} found other clouds near the western wing and a straight extension of the eastern wing in the velocity range 73 -- 84 km s$^{-1}$. Thus, the intrinsic velocity range of the ISM interacting with W50 has not been identified yet. Because of the velocity variation of the candidate ISM, it is difficult to determine the distance of W50 more precisely than as being between 3.0 and 5.5 kpc. Determining the intrinsic velocity is also important for identifying the origin of the gamma-ray emission induced by the interaction with cosmic-ray protons and estimating their total energy.

In this paper, our purpose is to investigate the influence of the SS433 jet on the morphology of W50. We further aim to specify the velocity range of the interacting ISM. Therefore, we compare the radio continuum images over 33 years, including the latest observation with the NSF's Karl G. Jansky Very Large Array (VLA) in the L band (1.0 -- 2.0 GHz) in 2017. We also determine the distribution of the HI gas around W50 using the archived survey data. From these results, we estimate the maximum energy of accelerated cosmic-ray particles in the SS433 jet terminal. Finally, we estimate the total energy of cosmic-ray protons associated with the GeV gamma-ray emission, assuming interaction with the HI gas. This paper is structured as follows: In Section 2, we describe the observational data and the method of calibration; in Section 3, we show the results of our analysis; in Section 4, we provide the discussion; in Section 5, we provide our conclusion.

\section{Observation and Calibration}
\subsection{Radio Continuum}
\subsubsection{Observation in 2017}
W50 was observed with the VLA in the L band on 2017 April 25 (project code: 17A-291; PI: G. M. Dubner). The array was in the D configuration during the observations. W50 extends 1$^{\circ}$${\times}$2$^{\circ}$ in size; however, the primary beam size is only 28$^{\prime}$ at 1.5 GHz. Then, a 58-pointing mosaic was used to cover all of the structures. The pointing separation was 15$^{\prime}$. The pointings satisfied Nyquist sampling at 1.4 GHz. Each pointing was observed twice in snapshots of 152 s.

The data were calibrated and imaged using the Common Astronomy Software Applications ({\sc CASA}) package following standard procedures. J1331+3030 and J1859+1259 are the flux density calibrator and the phase calibrator, respectively. L-band observations are severely affected by radio frequency interference (RFI). We checked the frequency channel affected by RFI and flagged the data below 1.327 GHz, between 1.520 and 1.711 GHz, and above 1.904 GHz. 

Diffuse emission from W50 is hidden by significant artifacts because some bright sources such as SS433 exist within the observation field. To reduce the sidelobe level, we first carried out clean deconvolution of the image using task {\tt tclean} while masking only 15 bright point sources, including SS433. We repeated {\tt tclean} until all the peak fluxes reached five times the rms noise level. The artifacts seemed improved at this stage. After {\tt tclean} was performed on the bright point sources, we performed {\tt tclean} for other radio sources in the target field including W50. Because W50 has extended structures, we applied the Multi-Scale, Multi-Term Multi-Frequency Synthesis (MS-MT-MFS) algorithm setting nterms = 1 \citep{rau2011}. The phase center was at 19$^{\rm h}$ 12$^{\prime}$ 17$^{\prime\prime}$, 4$^{\circ}$ 58$^{\prime}$ 00$^{\prime\prime}$. Images were made using Briggs weighting with robust = 0.5. We also set a low loop gain, 0.005, in order not to miss diffuse and faint emission. Finally, the image had a resolution of 53$^{\prime\prime}$${\times}$ 43$^{\prime\prime}$ with a position angle of -47$^{\circ}.$9 (0.8 pc ${\times}$ 0.6 pc at a distance of 3.0 kpc and 1.4 pc ${\times}$ 1.1 pc at a distance of 5.5 kpc).

In addition to the full-frequency image, we made an image whose frequency range accords with the observations in 1984 and 1996 to compare the structures in detail. The image was constructed through the same process as for the full-frequency image. We smoothed the final image with a beam size of 60$^{\prime\prime}$${\times}$ 60$^{\prime\prime}$ (0.9 pc at a distance of 3.0 kpc and 1.6 pc at a distance of 5.5 kpc) using task {\tt imsmooth} to equalize the beam size between all observation periods.

\subsubsection{Calibration and Imaging of Past Observation Dataset}
To examine variations of structures of W50 over 33 years, we also calibrated and imaged the observation data in 1984 and 1996. W50 was observed with the VLA at 1.465 GHz on 1984 August 12 and 13 \citep{elston1987}. The array configuration was also D, and 9-pointings were used. W50 was also observed with the VLA at 1.465 GHz on 1996 August 19 \citep{dubner1998}. The array configuration was also D, and mosaicking was carried out with 58-pointings. We calibrated this dataset with {\sc CASA} following standard procedures. The image was constructed through the same process as for the full-frequency image of the 2017 observation. We smoothed the final images with a convolved beam size of 60$^{\prime\prime}$${\times}$ 60$^{\prime\prime}$ using task {\tt imsmooth}.

\subsection{The GALFA-HI Survey}
To search the interstellar medium associated with W50, we used the Galactic Arecibo L-band Feed Array HI (GALFA-HI) survey data \citep{peek2011}. The survey was carried out with the Arecibo Observatory 305-m telescope. Its angular resolution is 4 arcmin, and the velocity resolution is 0.18 km s$^{-1}$. The velocity coverage is -700 km s$^{-1}$ $<$ V$_{\rm LSR}$ $<$ +700 km s$^{-1}$. Its typical noise level is 80 mK rms in an integrated 1 km s$^{-1}$ channel.

\section{Results}
\subsection{Comparison of the Structures of W50 during 33 Years}
\begin{figure*}[t]
\begin{center}
	\includegraphics[width=16cm,bb=0 0 1850 650]{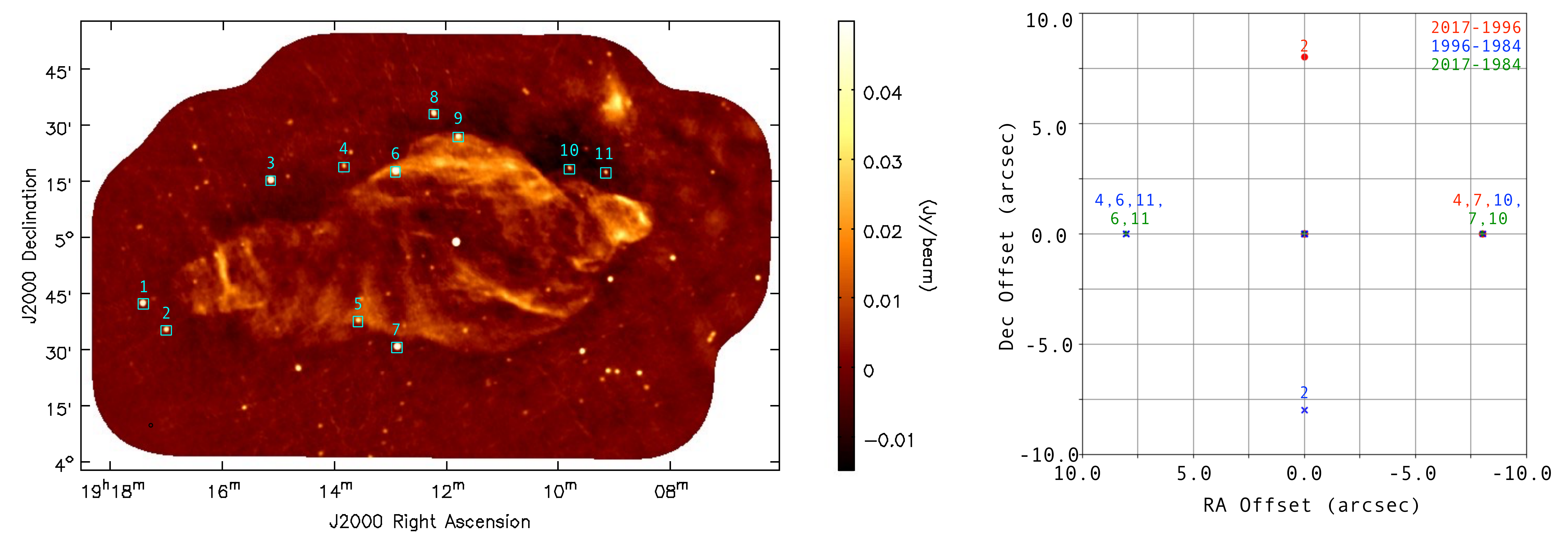}
    \caption{(Left) Total intensity map of W50 observed in 2017. Cyan boxes show the positions of bright radio sources used to assess the accuracy of position determination. (Right) Offsets of the positions of bright radio sources between observation periods. Comparisons between 2017 and 1996, 1996 and 1984, and 2017 and 1984 are represented with a red circle, blue cross, and green plus, respectively. We label only the sources that have offsets.}
    \label{fig:figureA1}
\end{center}
\end{figure*}

\subsubsection{Astrometric Error}
\label{sec:astrometric_error}
\begin{table}
 \tbl{Position offset of the flux calibrator.\footnotemark[$*$]}{%
\begin{tabular}{lcc}
 \hline
  \multicolumn{1}{c}{Year} & Right Ascension (arcsec) & Declination (arcsec) \\
 \hline
  1984 & +8.74524 & -7.55885 \\
  1996 & -0.01476 & -0.01885 \\
  2017 & -0.01476 & +0.00115 \\
 \hline
\end{tabular}}\label{tab:offset3C286}
\begin{tabnote}
 \footnotemark[$*$] Comparison with the position of 3C286 on the list of VLA calibrators. The position is RA (J2000.0): 13$^{\rm h}$31$^{\rm m}$08$^{\rm s}$.287984, Dec (J2000.0): +30$^{\circ}$30$^{\prime}$32$^{\prime \prime}$.958850.
\end{tabnote}
\end{table}

To discuss detailed motions using multiple-period observations, it is important to determine the astrometric error. We used the alignment accuracy of point sources that did not relocate during the observation interval. We first checked offsets in the position of the flux calibrator 3C286 by comparing the reference position that the observatory offers. 3C286 was imaged using the {\sc CASA} task {\tt tclean}. After {\tt tclean}, we smoothed with a convolved beam size of 60$^{\prime\prime}$${\times}$ 60$^{\prime\prime}$ using task {\tt imsmooth}. We fitted a Gaussian function to determine the position of 3C286. Finally, we obtained the offset values by determining the difference from the reference position. The results are listed in Table \ref{tab:offset3C286}. The position accuracy in 1984 is poor compared with the 1996 and 2017 observations, and the cause might be the weather or imperfect correction of the antenna position.

We also checked the positions of bright radio sources in the target field. The sources are shown in the left panel of Figure \ref{fig:figureA1} with cyan boxes. We calculated the position offsets of these sources between 2017 and 1996 (red), 1996 and 1984 (blue), and 2017 and 1984 (green) (right panel of Figure \ref{fig:figureA1}). Most sources have no offset; however, the positions of some bright sources are slightly different between observation periods. The numbers in the plot correspond to the source numbers in the left panel. Their offset values coincide with the pixel size of the total intensity maps, $\sim$ 8 arcsec. 

Discussing the motion of a radio component on a small scale requires caution and sometimes causes misunderstanding. Therefore, we adopt the largest offset of 3C286 in 1984, 8.74524 arcsec, as the astrometric error.


\subsubsection{Comparison of Total Intensity Distributions}
\begin{figure}[th]
\begin{center}
	\includegraphics[width=8cm,bb=0 0 777.25 468.14]{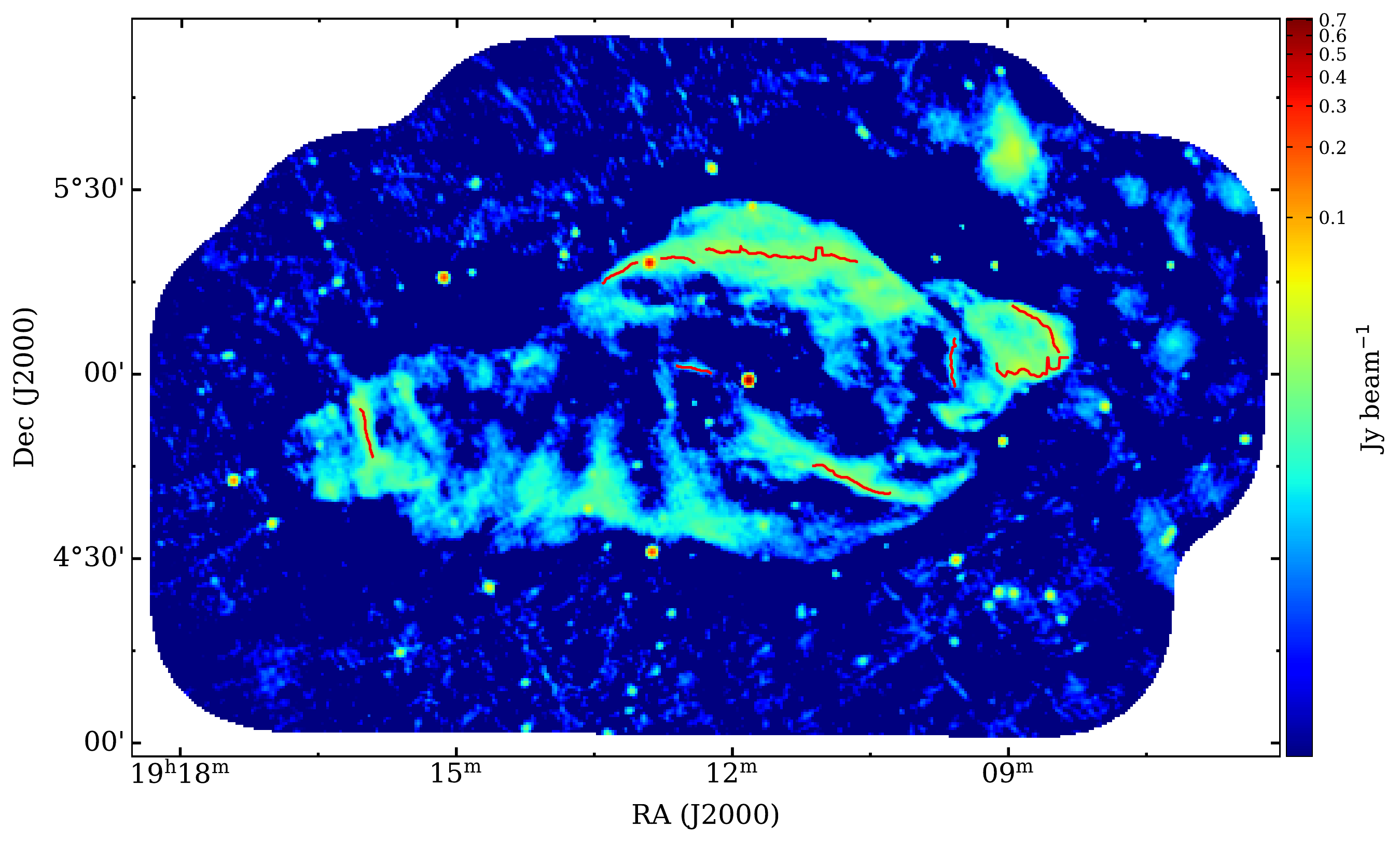}
    \caption{Positions of the filamentary structures observed in 2017. The back image is a total intensity map at 1.440 -- 1.490 GHz. The beam size is 60$^{\prime\prime}$${\times}$ 60$^{\prime\prime}$.}
    \label{fig:figureA2_2017}
    \end{center}
\end{figure}

\begin{figure}[th]
\begin{center}
	\includegraphics[width=8cm,bb=0 0 777.25 468.14]{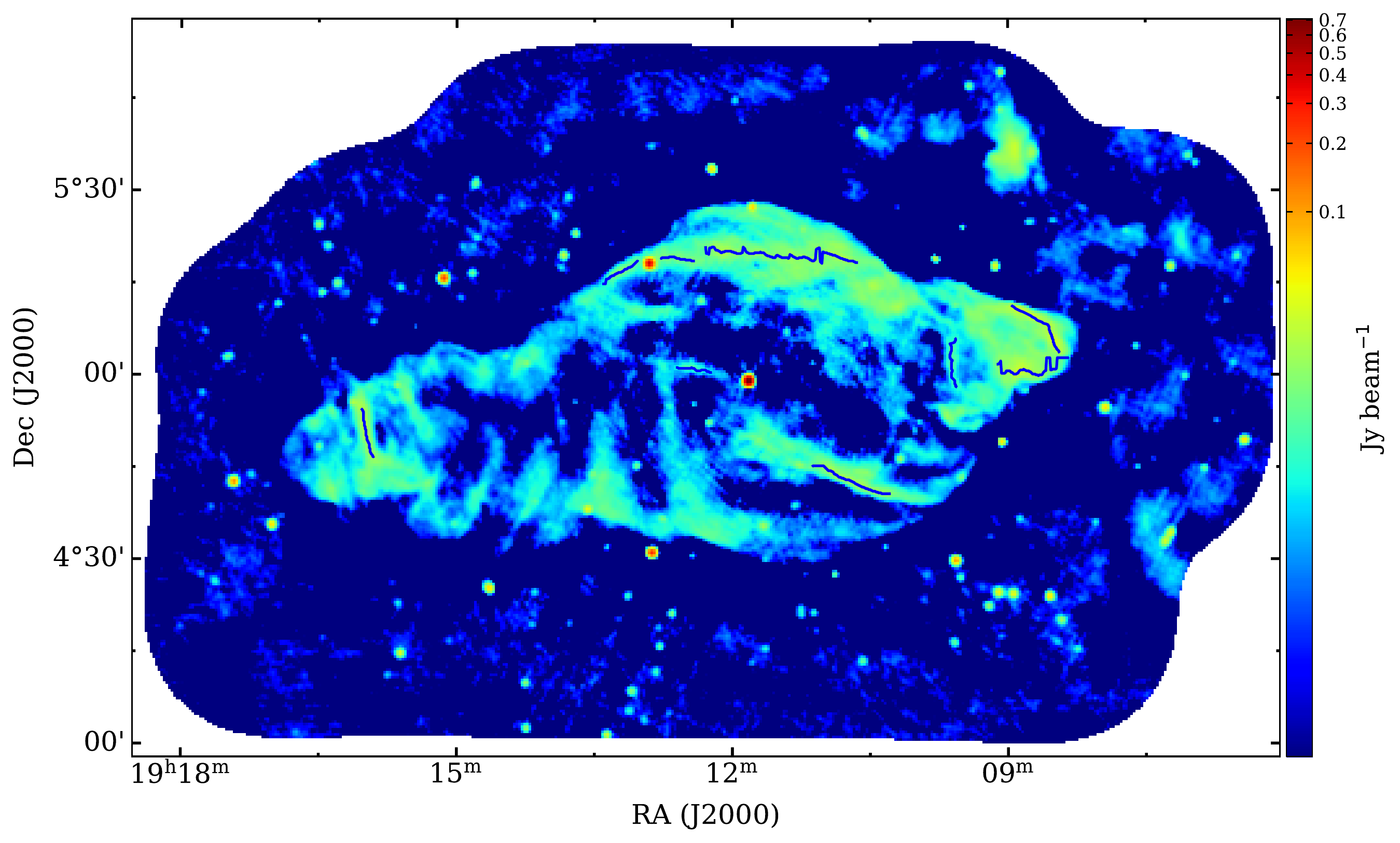}
    \caption{Similar to Figure \ref{fig:figureA2_2017}, but for the observation in 1996.}
    \label{fig:figureA2_1996}
    \end{center}
\end{figure}

\begin{figure}[th]
\begin{center}
	\includegraphics[width=8cm,bb=0 0 777.25 468.14]{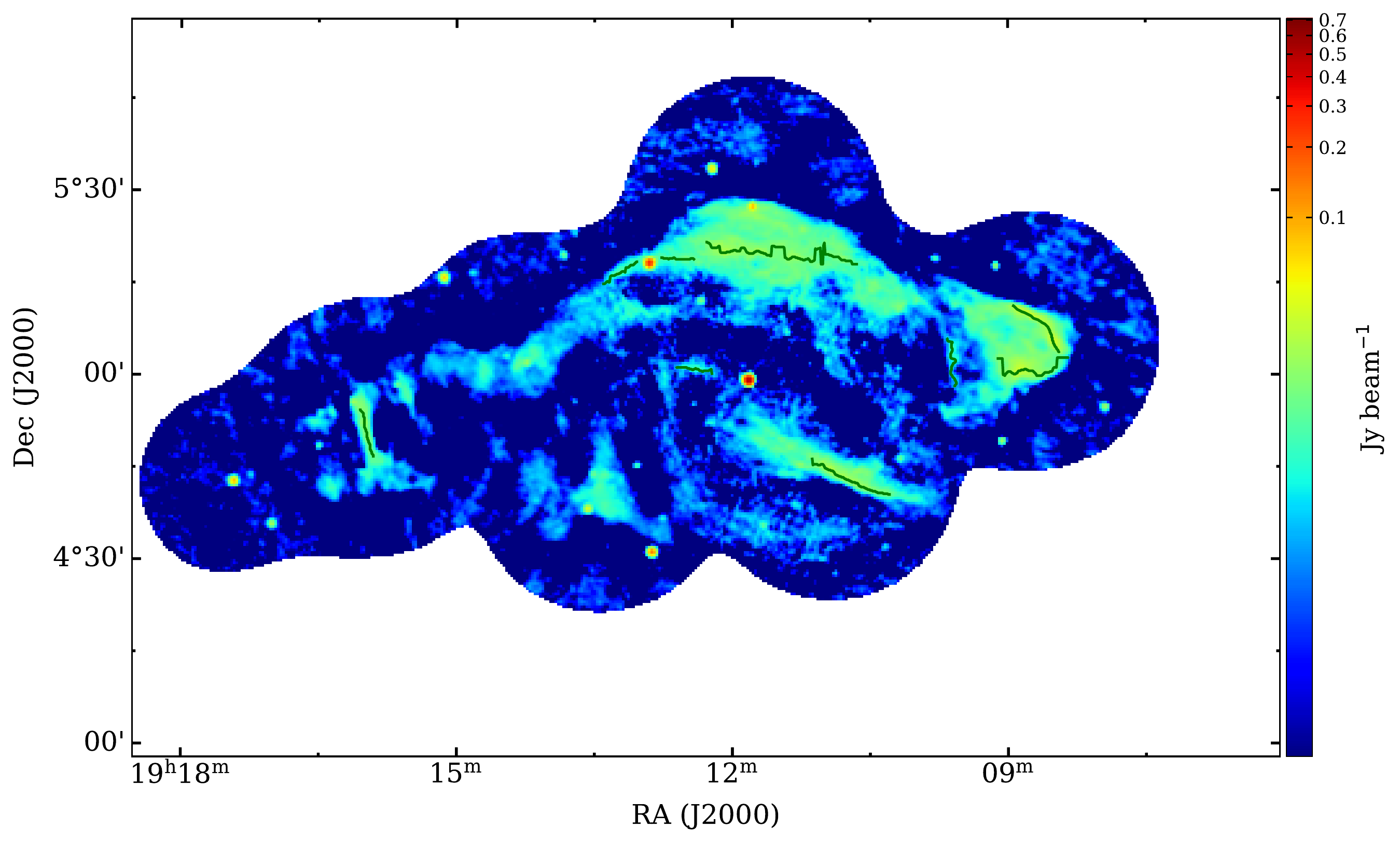}
    \caption{Similar to Figure \ref{fig:figureA2_2017}, but for the observation in 1984.}
    \label{fig:figureA2_1984}
    \end{center}
\end{figure}

\begin{figure*}[t]
\begin{center}
	\includegraphics[width=16cm,bb=0 0 1200 500]{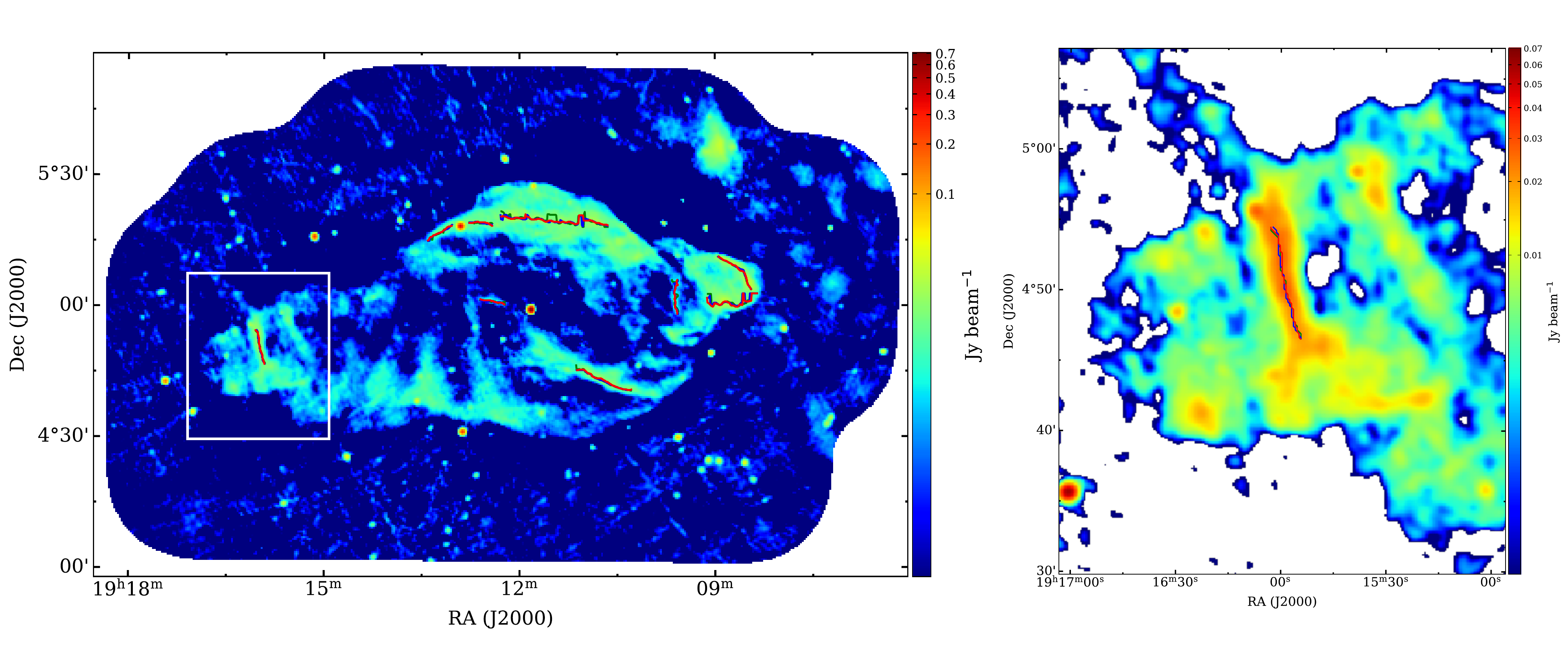}
    \caption{(Left) Comparison of the peak positions of the bright filaments associated with W50. The green, blue, and red lines represent the peak positions of the filaments observed in 1984, 1996, and 2017, respectively. The background image shows the total intensity distribution between 1.440 and 1.490 GHz observed in 2017. The beam size is 60$^{\prime\prime}$${\times}$ 60$^{\prime\prime}$. The white box shows the position of the ETF (right panel).}
    \label{fig:figure2}
    \end{center}
\end{figure*}

To determine whether the morphology of W50 changed over 33 years, we compared the observations in 1984, 1996, and 2017. We made the total intensity map of each observation period using the same {\tt tclean} setup: phase reference, map size, grid size, and weighting (Figures \ref{fig:figureA2_2017}, \ref{fig:figureA2_1996}, and \ref{fig:figureA2_1984}). Furthermore, we equalized the beam size to be 60$^{\prime\prime}$${\times}$ 60$^{\prime\prime}$. After that, we found the peak positions of the bright filaments.

The green, blue, and red curves in Figure \ref{fig:figure2} represent the peak positions of the characteristic filaments observed in 1984, 1996, and 2017, respectively. The bright filaments do not show significant motions above the astrometric error during the 33 years. Remarkably, the filament perpendicular to the jet axis, the ETF, seems to remain stationary even though it should be influenced by the SS433 jets (right panel of Figure \ref{fig:figure2}). We can use this observation to estimate the speed of the ETF (see Section. 4.1).

\subsection{Neutral Hydrogen around W50}
\label{sec:HI}
\begin{figure*}[htbp]
\begin{center}
	\includegraphics[width=16cm,bb=0 0 1000 1050]{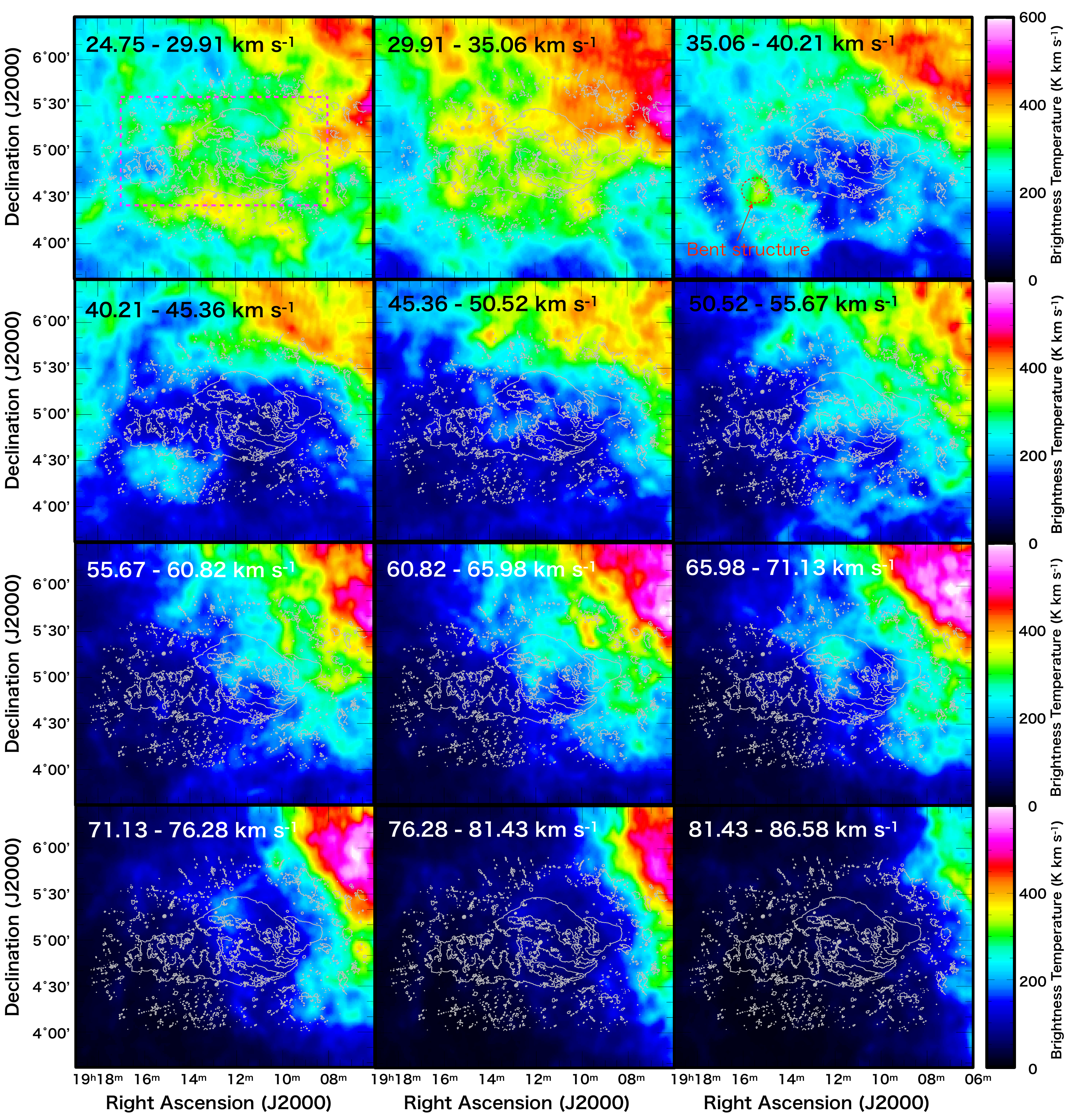}
    \caption{Channel map of HI line emission observed with the GALFA-HI survey. The velocity range is from 24.75 km s$^{-1}$ to 86.58 km s$^{-1}$ with a velocity bin of 5.15 km s$^{-1}$. The angular resolution is 4 arcmin. The grey contours show the total intensity between 1.440 and 1.490 GHz observed with the VLA in 2017. The magenta box in the top-left panel indicates the region referred to in Figure 4.}
    \label{fig:figure3}
\end{center}
\end{figure*}

\begin{figure*}[htbp]
\begin{center}
	\includegraphics[width=16cm,bb=0 0 1000 580]{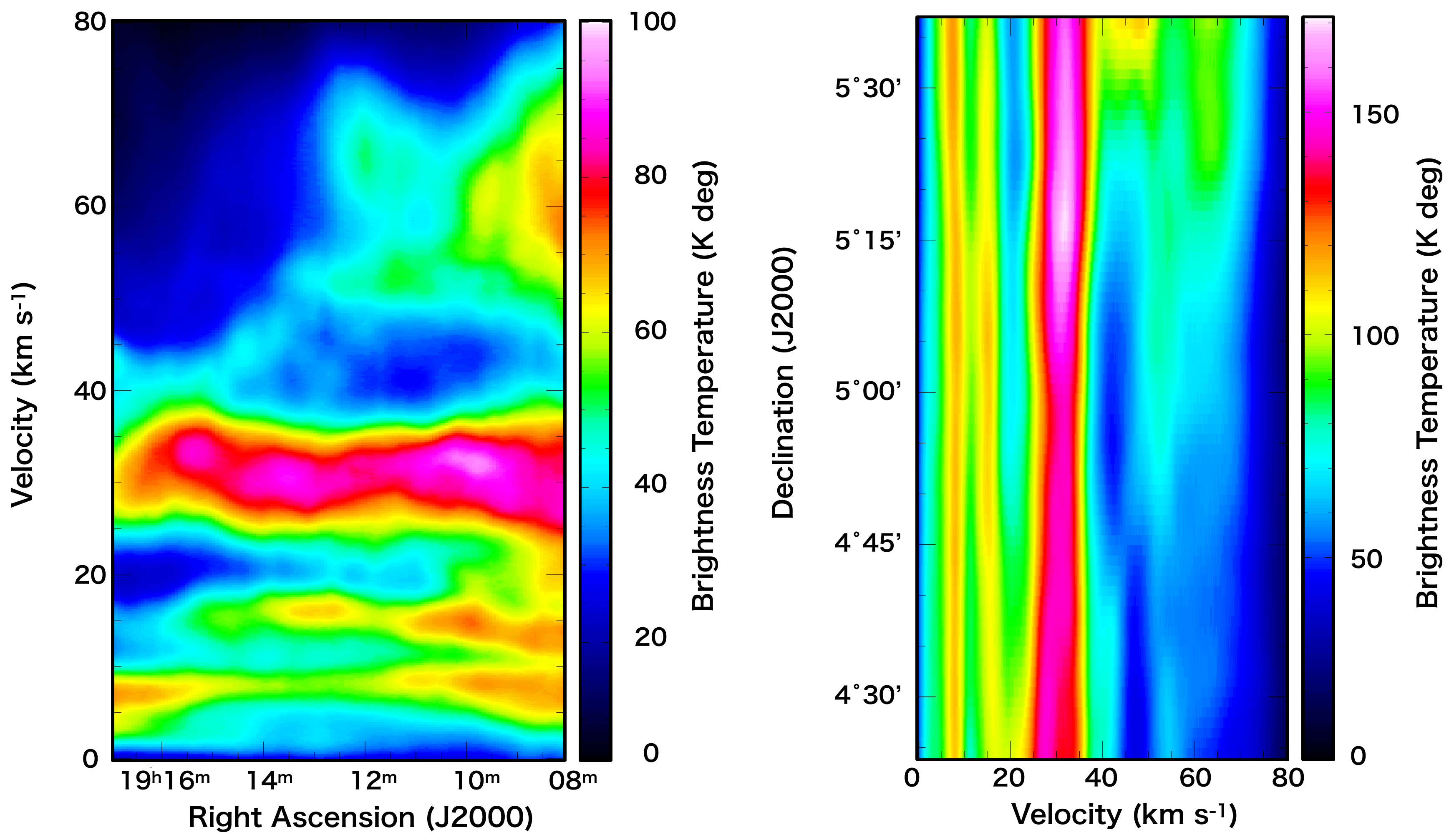}
    \caption{Position-velocity diagrams around W50. The integrated region is shown in Figure \ref{fig:figure3}. The left panel is the velocity distribution in the direction of right ascension. The right panel is a similar map for the direction of declination.}
    \label{fig:figure4}
\end{center}
\end{figure*}

Figure \ref{fig:figure3} shows a channel map of HI emission surrounding W50. The velocity range is from 24.75 km s$^{-1}$ to 86.58 km s$^{-1}$, and the channel bin is 5.15 km s$^{-1}$. The HI cavity found in the velocity range from 35.06 km s$^{-1}$ to 50.52 km s$^{-1}$ appears to coincide spatially with W50 and has been identified by \citet{dubner1998}. Furthermore, in the velocity range of 35.06 km s$^{-1}$ to 40.21 km s$^{-1}$, a dense cloud is at the bent structure of the eastern wing. The HI cloud at the bent structure goes out above 45.36 km s$^{-1}$, and emission associated with the Galactic plane gradually departs from W50. Above 71.13 km s$^{-1}$, the HI cloud associated with the Galactic plane is parallel to the edge of the western wing. \citet{su2018} detected a molecular cloud in this area, and the velocity range also coincided well.

Both panels of Figure \ref{fig:figure4} are position-velocity diagrams. The dashed magenta box in the upper-left panel of Figure \ref{fig:figure3} shows the integrated area of these position-velocity diagrams. A cavity-like structure appears in both diagrams centered on the velocity 44 km s$^{-1}$ with a range between 33 km s$^{-1}$ and 55 km s$^{-1}$. These trends correspond to the cavity surrounding W50 shown in Figure \ref{fig:figure3}. A similar structure is seen in SNRs \citep{sano2017,sano2019}. An SNR sweeps up surrounding HI clouds, and a cavity appears in an intensity map and a position-velocity diagram. Cavity structures also can be observed around AGNs as a result of interaction between outflows and the surrounding medium \citep{panagoulia2014}. This means that W50 can form a cavity regardless of the origin of the spherical shell of W50, whether an SNR or jets from SS433. Therefore, the HI cavity is evidence of the interaction between W50 and the interstellar medium. Assuming the velocity field of our galaxy, we can derive the distance of W50 using the velocity information of the related HI gas \citep{brand1993}. The distance of W50 is estimated as 3.0 kpc using the center velocity of the cavity, 44 km s$^{-1}$. As mentioned in Section \ref{sec:intro}, this distance is different from values obtained with other methods. The distance of W50 is still under discussion, and it is difficult to conclude from our analysis. We mention the distance of the SS433/W50 system again in Section \ref{sec:jet model}.



\section{Discussion}
In this section, we estimate the velocity of the ETF (Filament 5 in Figure \ref{fig:figure1} in the Appendix). We also construct a jet model and discuss the deceleration of the jet. Using the velocity estimate, we derive the maximum energy of cosmic-ray particles accelerated at the jet terminal. Finally, we consider the GeV gamma-ray emission scenario of interaction with the HI cavity and the total energy of protons accelerated by W50.
\begin{figure*}[t]
\begin{center}
	\includegraphics[width=8cm,bb=0 0 700 700]{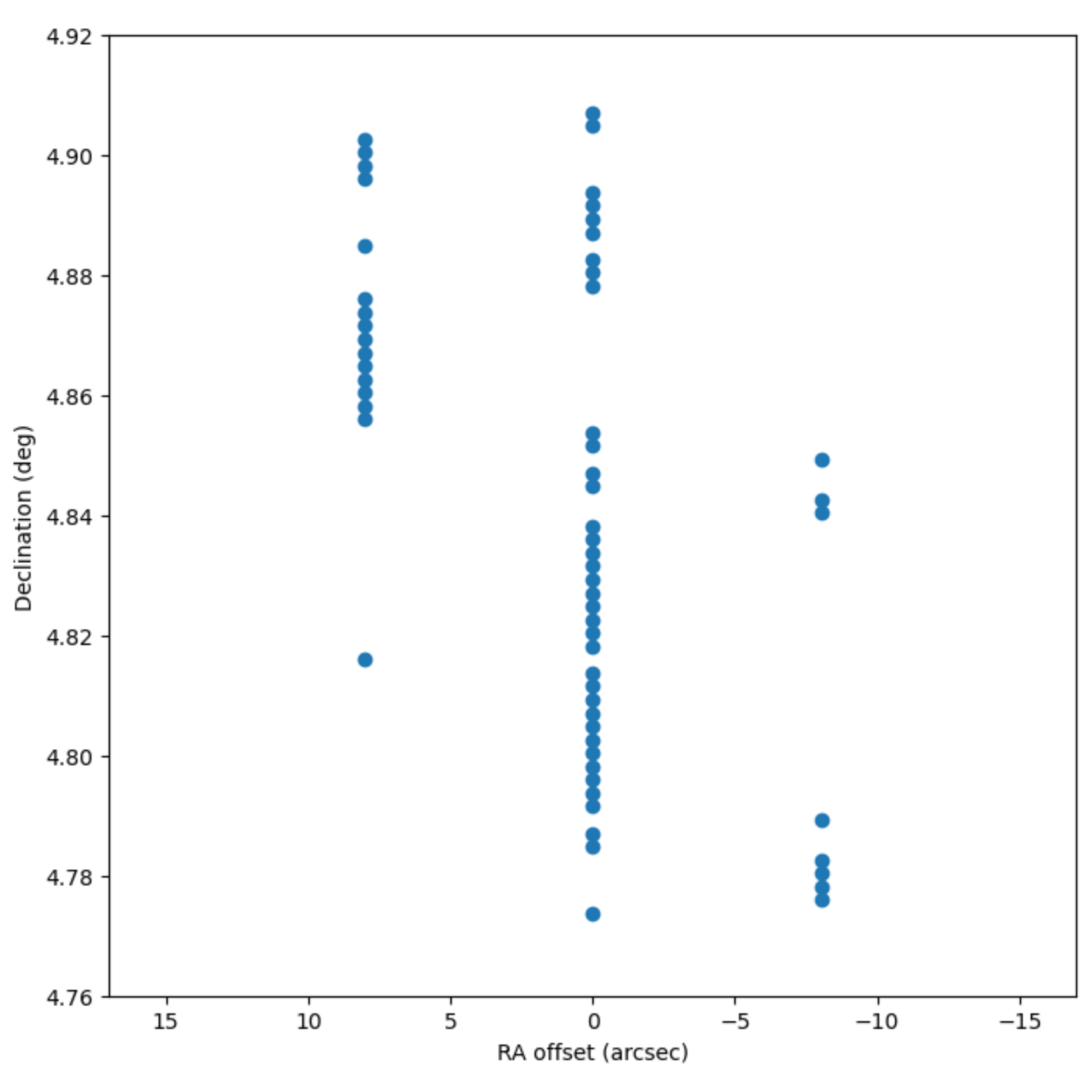}
    \caption{Right ascension offsets of the peak positions along the ETF from SS433 between 2017 and 1984.}
    \label{fig:filament-offset}
\end{center}
\end{figure*}

\subsection{Velocity of the Eastern Terminal Filament (ETF)}
\label{sec:upperlimit_vel}
\citet{goodall2011a} derived the upper limit velocity of the ETF as 0.0405$c$. This means that for a propagation velocity of 0.0405$c$, the filament observed in 2017 should shift eastward by 15 arcsec from the position in 1984, assuming a distance of 5.5 kpc. If we assume a distance of 3.0 kpc, the upper limit velocity becomes 0.0221$c$, and the filament also moves about 15 arcsec. The expected shift is larger than the astrometric error. We compared the positions of the ETF between 1984 and 2017 as follows.

First, we examined the right ascension offsets between SS433 and the peak positions of the filament in each observation period. After that, we subtracted the offsets of the 1984 observation from those of the 2017 observation. The result is shown in Figure \ref{fig:filament-offset}. None of the peak positions of the ETF has an offset exceeding the astrometric error of $\sim$ 8.74524 arcsec. This means that the ETF was stable for 33 years and did not move with the upper limit velocity estimated by \citet{goodall2011a}.

Using the above result, we can estimate a new upper limit velocity. We cannot distinguish a shift below the astrometric error $\delta_{\rm R.A}$ = 8.74524 arcsec. This corresponds to 0.13 pc = 0.41 ly assuming a distance of 3.0 kpc and to 0.23 pc = 0.76 ly when we assume a distance of 5.5 kpc. Accordingly, we can estimate the upper limit velocity of the ETF as
\begin{equation}
	v^{\rm 3.0kpc}_{\rm 33yr}=\frac{0.41}{t_{\rm base}}=0.013\ c\sim 3,897\ {\rm km\ s^{-1}},
	\label{eq:upvel33_2}
\end{equation}
for a distance of 3.0 kpc and
\begin{equation}
	v^{\rm 5.5kpc}_{\rm 33yr}=\frac{0.76}{t_{\rm base}}=0.023\ c\sim 6,895\ {\rm km\ s^{-1}},
	\label{eq:upvel33_1}
\end{equation}
for 5.5 kpc, where $t_{\rm base}$ = 32.70 yr on the basis of a 365.25-d year. We investigate the detectability of the ETF motion with a future radio telescope such as the Square Kilometre Array (SKA) Phase 1 and the Next Generation Very Large Array (ngVLA), assuming that the ETF moves with either of these upper limit velocities. On the basis of the predicted performances of these telescopes, we conclude that a 6-year observation is required to identify the motion\footnote{ngVLA Memo Series No.17 $\langle$http://library.nrao.edu/ngvla.shtml$\rangle$}\citep{braun2019}.


\subsection{Comparison between Observation and Energy Conservation Jet Model}
\label{sec:jet model}
\begin{figure*}[t]
\begin{center}
	\includegraphics[width=16cm,bb=0 0 879 346]{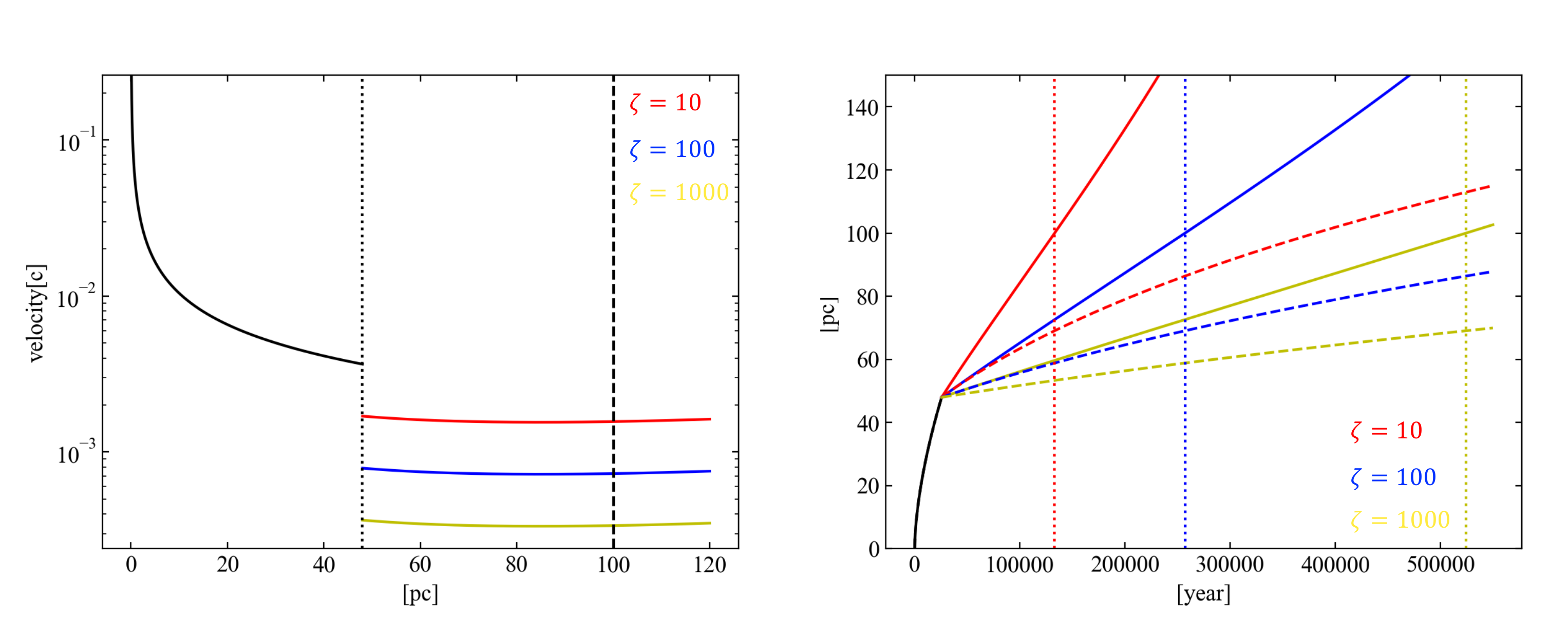}
    \caption{(Left) Relation between the distance from SS433 and velocity of the eastern terminal in Model A. The distance of SS433 is assumed to be 5.5 kpc. The black solid line shows the relation within the spherical shell, whose radius is marked with a black dotted line ($|x_1|$ = 48 pc). The red, blue, and yellow solid lines correspond to the relation outside the shell for $\zeta$ = 10, 100, and 1000, respectively. The black dashed line shows the recent position of the eastern terminal of the jet ($x_t$ = 100 pc). (Right) Relation between the time and distance of the eastern and western terminals in Model A. The solid and dashed lines correspond to the eastern and western terminals, respectively. The dotted lines show the time when the eastern terminal region reaches the current position.}
    \label{fig:w-shell}
\end{center}
\end{figure*}

\begin{figure*}[t]
\begin{center}
	\includegraphics[width=16cm,bb=0 0 879 346]{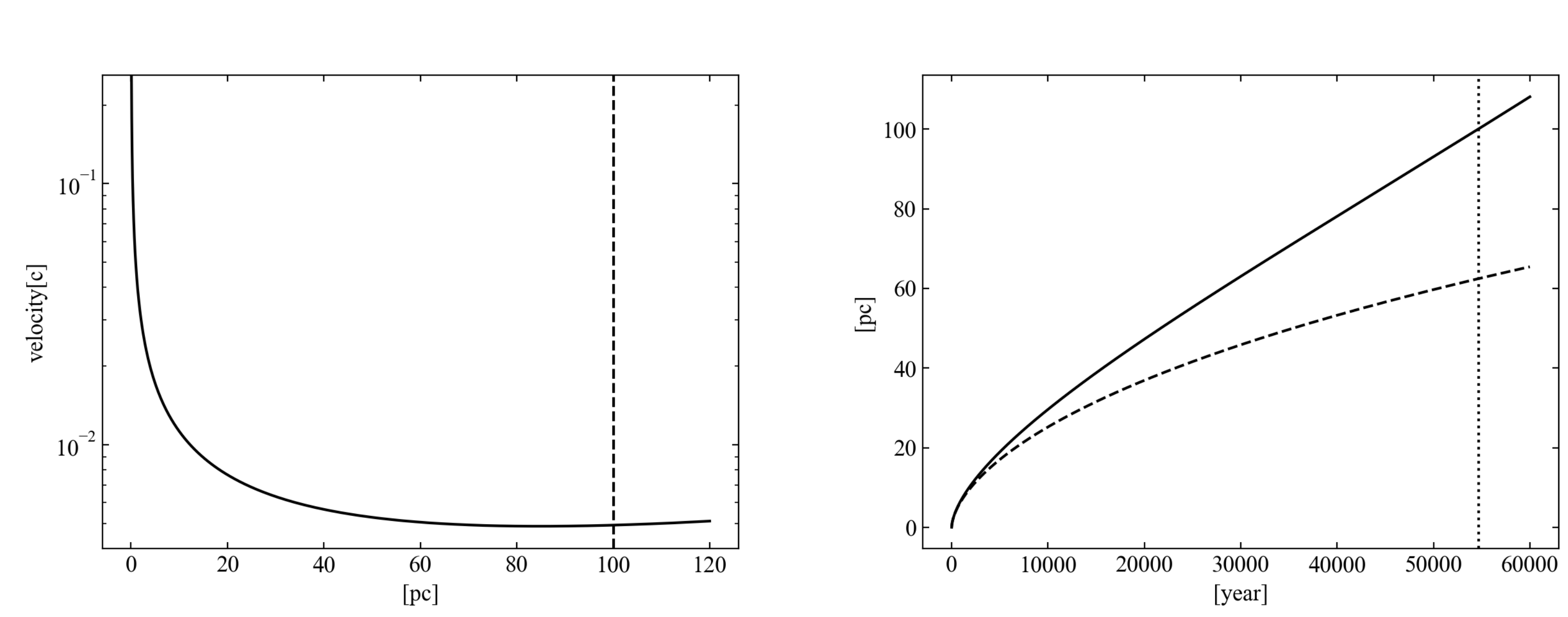}
    \caption{Similar to Figure \ref{fig:w-shell}, but for Model B.}
    \label{fig:no-shell}
\end{center}
\end{figure*}

\citet{goodall2011a} suggested that a continuous, non-multi-episodic jet cannot be decelerated below the upper limit velocity 0.0405 $c$ by interaction with the surrounding ISM. However, \citet{panferov2017} suggested that the jet could be decelerated by viscosity between the jet and its surroundings. They also estimated the jet velocity decreases by $\sim60\%$ within the spherical shell of W50, although the value is still higher than the estimated upper limit velocity in Section \ref{sec:upperlimit_vel}.

Here, we use a one-dimensional model of the velocity of the terminal of the continuous jet assuming kinetic energy flux conservation \citep{zaninetti2016} with and without the initial spherical shell of W50. On the basis of the result, we determine whether the continuous jet can be decelerated below the upper limit velocity at the current position of the eastern jet terminal. We also discuss the formation scenario of W50 considering the formation timescale and the propagation distance of the western jet.

We define $x$ as the distance from SS433 along the jet axis on the pc scale, and the east of SS433 is the positive direction. The jet velocity around SS433 was estimated as $v_{0}$ = 0.26 $c$ at a distance of $x_{0}$ = 0.04 and 0.08 pc from SS433 assuming distances of 3.0 and 5.5 kpc, respectively \citep{hjellming1981,blundell2004}. The kinetic energy flux at $x_{0}$ should be conserved at $x$. The cross-sectional area of the cone formed by the SS433 helical jet, $A(x)$, is expressed as
\begin{equation}
	A(x)=\pi \left(x \tan\alpha\right)^{2},
	\label{eq:secarea}
\end{equation}
where $\alpha$ = 20$^{\circ}$ is the opening angle of the cone \citep{margon1984}. We define the radial distance of SS433 from the Galactic center as $R_{\rm SS433}$ and the height from the Galactic disk as $z_{\rm SS433}$. When the SS433 jet reaches $x$, the radial distance $R$ and the height $z$ of the jet terminal are written as
\begin{equation}
R=x\sin{\beta}+R_{\rm SS433},
\end{equation}
\begin{equation}
z=x\cos{\beta}+z_{\rm SS433},
\end{equation}
where $\beta$ = 19.7$^{\circ}$ is the inclination of the jet axis toward the normal line of the Galactic disk \citep{goodall2011b}.

The SS433/W50 system is located near the Galactic disk, and the surrounding ISM has a Galactic exponential profile \citep{dehnen1998},
\begin{equation}
\rho(R,z)=\rho_0\exp{\left[-\frac{R_m}{R_d}-\frac{R}{R_d}-\frac{z}{z_d}\right]},
\label{eq:density}
\end{equation}
where $\rho_{0}$ is the prefactor density determined from a normalization condition at the location of SS433, $R_m = 4\ {\rm kpc}$ is the constant length of the Galactic ISM disk, $R_d = 5.4\ {\rm kpc}$ is the scale length, and $z_{d}$ = 40 pc is the scale height from the Galactic disk. We consider two models on the basis of this ISM profile. First, we assume a formation scenario of W50 with the spherical shell (by wind bubble or SNR) + jet, hereafter called Model A. In this model, there is a spherical shell with the initial radius $|x_{1}|$, and a two-sided jet propagates within the shell where the density $\rho_{\rm shell}$ is constant. After the jet breaks through the shell, it proceeds through the surrounding ISM, which has the exponential density gradient in equation \ref{eq:density}. Then, the kinetic energy flux conservation of the jet can be expressed as
\begin{equation}
	\frac{1}{2} \rho_{\rm shell} v_{0}^{3} A(x_{0})=\frac{1}{2} \rho_{\rm shell} v(x)^{3} A(x)\ (|x|\leq |x_{1}|),
	\label{eq:enepres1}
\end{equation}
\begin{eqnarray}
	&\frac{1}{2}& \rho_{\rm 0} \exp\left[-\frac{R_m}{R_d}-\frac{x_1\sin{\beta}+R_{\rm SS433}}{R_d}-\frac{x_1\cos{\beta}+z_{\rm SS433}}{z_{d}}\right] V(x_{1})^{3} A(x_{1}) \nonumber \\
	=&\frac{1}{2}& \rho_{\rm 0} \exp\left[-\frac{R_m}{R_d}-\frac{x\sin{\beta}+R_{\rm SS433}}{R_d}-\frac{x\cos{\beta}+z_{\rm SS433}}{z_{d}}\right] V(x)^{3} A(x)\ (|x|>|x_{1}|),
	\label{eq:enepres2}
\end{eqnarray}
where $v(x)$ and $V(x)$ are the velocities of the terminal in the ranges $|x|\leq |x_{1}|$ and $|x|>|x_{1}|$, respectively. Because the kinetic energy flux is conserved inside the spherical shell, the following relation is satisfied:
\begin{equation}
\left[\frac{v(x_1)}{V(x_1)}\right]^{3}=\frac{\rho_0}{\rho_{\rm shell}}\exp{\left[-\frac{R_m}{R_d}-\frac{x_1\sin{\beta}+R_{\rm SS433}}{R_d}-\frac{x_1\cos{\beta}+z_{\rm SS433}}{z_d}\right]}\equiv\zeta.
\end{equation}
$\zeta$ corresponds to the density ratio of the surrounding ISM at the boundary and within the spherical shell. Therefore, the velocity at $x$ is
\begin{equation}
	v(x)=v_{0}\left(\frac{x_{0}}{x}\right)^{\frac{2}{3}}\ (|x|\leq |x_{1}|),
	\label{eq:velocity1}
\end{equation}
\begin{equation}
	V(x)=v(x_{1})\zeta^{-\frac{1}{3}}\left(\frac{x_{1}}{x}\right)^{\frac{2}{3}}\left\{\exp \left[\frac{(x-x_{1})\sin{\beta}}{R_{d}}+\frac{(x-x_{1})\cos{\beta}}{z_{d}}\right]\right\}^{\frac{1}{3}}\ (|x|>|x_{1}|).
	\label{eq:velocity2}
\end{equation}
As shown in equations \ref{eq:velocity1} and \ref{eq:velocity2}, the jet velocity changes discontinuously at the boundary. We may interpret the velocity jump as the dissipation by the shock wave. The typical density within an SNR shell is believed to be in the range 0.01 -- 1 ${\rm cm}^{-3}$, and the density of the surrounding ISM of SS433 should be higher because SS433 is located near the Galactic disk. We consider the cases $\zeta = 10, 100,$ and $1000$.

Second, we consider a no-initial-spherical-shell scenario, hereafter called Model B. This means that the SS433 jet propagates in the surrounding ISM and reaches the current terminal position without the initial spherical shell. In this case, the kinetic energy flux conservation of the jet can be expressed as
 \begin{eqnarray}
	&\frac{1}{2}& \rho_{\rm 0} \exp\left[-\frac{R_m}{R_d}-\frac{x_0\sin{\beta}+R_{\rm SS433}}{R_d}-\frac{x_0\cos{\beta}+z_{\rm SS433}}{z_{d}}\right] v_0^{3} A(x_0)\\
	=&\frac{1}{2}& \rho_{\rm 0} \exp\left[-\frac{R_m}{R_d}-\frac{x\sin{\beta}+R_{\rm SS433}}{R_d}-\frac{x\cos{\beta}+z_{\rm SS433}}{z_{d}}\right] v(x)^{3} A(x),
	\label{eq:enepres3}
\end{eqnarray}
and the velocity at $x$ is 
\begin{equation}
	v(x)=v_0\left(\frac{x_{0}}{x}\right)^{\frac{2}{3}}\left\{\exp \left[\frac{(x-x_{0})\sin{\beta}}{R_{d}}+\frac{(x-x_{0})\cos{\beta}}{z_{d}}\right]\right\}^{\frac{1}{3}}.
	\label{eq:velocity3}
\end{equation}

Figure \ref{fig:w-shell} (left) shows the relation between the velocity of the terminal of the eastern jet and distance from SS433 in Model A (equations \ref{eq:velocity1} and \ref{eq:velocity2}) assuming a distance of 5.5 kpc for SS433. In this case, the shell radius is $|x_1|$ = 48 pc and the terminal of the eastern jet propagates about $x$ = $x_t$ = 100 pc, where $x_t$ is the distance of the eastern terminal from SS433. At position $x_t$, the velocities of the terminal decrease to 0.0016$c$ (480 km s$^{-1}$), 0.00073$c$ (219 km s$^{-1}$), and 0.00034$c$ (102 km s$^{-1}$) for $\zeta = 10, 100,$ and $1000$, respectively. These values are all below the estimated upper limit velocity of 0.023$c$ (6,895 km s$^{-1}$). If we assume a distance of 3.0 kpc for SS433, $|x_1|$ = 26 pc, and $x$ = $x_t$ = 55 pc, the velocities of the terminal region are 0.0012$c$ (360 km s$^{-1}$), 0.00057$c$ (171 km s$^{-1}$), and 0.00026$c$ (77.9 km s$^{-1}$) for $\zeta = 10, 100,$ and $1000$, respectively. These results seem consistent with the observation. Figure \ref{fig:w-shell} (right) shows the relation between the time and distance from SS433. The solid and dashed lines correspond to the eastern and western jets, respectively. Because the distance of the terminal of the eastern jet is about $x_t$ = 100 pc from SS433, it takes 133,140 years, 257,192 years, and 524,454 years to reach the current position for $\zeta = 10, 100,$ and $1000$, respectively. These times are too long to maintain the continued jet. In addition, the western terminal propagates about 69 pc in all cases assuming a distance of 5.5 kpc, which is larger than the observed value of 56 pc by about $23 \%$. Thus, it is difficult to explain the formation of W50 with Model A.

Figure \ref{fig:no-shell} shows the corresponding results for Model B (equation \ref{eq:velocity3}). When the eastern jet propagates to the distance $x_t$ = 100 pc, the terminal velocity is 0.0049$c$ (1,469 km s$^{-1}$), and it takes 54,660 years to reach the current position. At that time, the terminal of the western jet propagates about 62 pc. Considering the propagation time scale and the distance ratio between the eastern and western wings, Model B is slightly better than Model A for a distance of 5.5 kpc. The $10 \%$ difference in the propagation scale of the western terminal between Model B and observation is still larger than the astrometric error. This could be explained by the local distribution of the surrounding ISM. Although we cannot conclude that Model B is more appropriate than Model A, the result implies that the SS433 jets might be able to construct the morphology of W50 without an initial wind-bubble shell or SNR, as suggested by \citet{ohmura2021}.

However, if we assume the distance of SS433 to be 3.0 kpc with Model B, the eastern terminal velocity is 0.0032$c$ (959 km s$^{-1}$), and it takes 39,392 years to reach the current position. At that time, the western terminal region propagates about 41 pc in this model. This is longer than the observed value of 31 pc. This result might imply that the distance of 3.0 kpc derived from the velocity of the HI cavity surrounding W50 is doubtful; however, we suggest that the HI cavity is still correlated with W50. Observational and numerical studies have revealed that the peculiar motions of gases, stars, and star-forming regions induced by localized turbulence could yield systematic errors in kinematic distances of a few kpc (see \cite{baba2009,ramon-fox2018} and references therein). Then, the central velocity of the HI cavity may reflect the influence of the peculiar motion of HI gas interacting with the SS433/W50 system.

\subsection{Maximum Energy of Accelerated Protons at the Jet Terminal Region}
In this section, we estimate the maximum energy of accelerated particles at the SS433 jet terminal. For reference, we derive the energy assuming both distances of 3.0 and 5.5 kpc. When the jet terminal shock induces diffusive shock acceleration (DSA) \cite{bell1978a,bell1978b,blandford1980,drury1983}, the maximum energy can be calculated with the equation
\begin{equation}
E_{\rm max}=6\times10^{14}\ {\rm eV}\frac{Z}{\eta_{B}}\left(\frac{v_s}{10^4\ {\rm km\ s^{-1}}}\right)\ \left(\frac{B}{10\ \mu{\rm G}}\right)\left(\frac{R}{2\ {\rm pc}}\right),
\label{eq:DSA}
\end{equation}
where $Z$ is the atomic number, $\eta_{B}$ is the degree of turbulence of the magnetic field, $v_{s}$ is the shock velocity, $B$ is the magnitude of the magnetic field, and $R$ is the scale size of the upstream. We consider the case of strong shock (Mach number $M\gg1$) and monochromatic gas (a heat capacity ratio $\gamma=5/3$). In this case, we obtain the velocity ratio $\xi=v_{1}/v_{2}=4$, where $v_{1}$ and $v_{2}$ are the velocities of the upstream and downstream, respectively. In the case of a strong shock, these velocities also have the relation $v_{1}=4v_{2}=v_{s}$.

We consider the acceleration of protons ($Z=1$) and the condition $\eta_{B}$ = 1. The magnitude of the magnetic field of the ETF can be estimated assuming equipartition between the total energy densities of cosmic-ray electrons and that of the magnetic field \citep{beck2005}. Assuming a constant ratio of the number densities of protons and electrons, ${\rm {\bf K_0}} = 100$, and a spectral index $\sim 0.82$ \citep{dubner1998}, we obtain the magnitudes $B^{3.0\ {\rm kpc}}=45\ \mu{\rm G}$ and $B^{5.5\ {\rm kpc}}=39\ \mu{\rm G}$. Note that these values are lower estimates because of the effect of the missing flux. Because now we consider the acceleration at the jet terminal shock, we set the distance between the ETF and the TeV gamma-ray emitter region as the scale size of the upstream, $R^{3.0\ {\rm kpc}}=24\ {\rm pc}$ and $R^{5.5\ {\rm kpc}}=44\ {\rm pc}$, where the other shock wave is formed. Assuming the downstream velocity to be the upper limit velocity of the ETF estimated in Section \ref{sec:upperlimit_vel}, we obtain the maximum energy of the accelerated protons at the jet terminal shock:
\begin{equation}
E^{3.0\ {\rm kpc}}_{\rm max}\sim5.1\times10^{16}\ {\rm eV},
\end{equation}
and
\begin{equation}
E^{5.5\ {\rm kpc}}_{\rm max}\sim1.4\times10^{17}\ {\rm eV}.
\end{equation}

We also estimate the lower and upper limits of the maximum energy. First, the lower limit of the maximum energy is obtained by adopting the speed of sound as the downstream velocity $v_2$. \citet{brinkmann2007} estimated the temperature of the eastern terminal of W50 to be $0.28\pm0.01\ {\rm keV}$. Then, the speed of sound is calculated to be 300 km s$^{-1}$. Assuming $v_{2}=300\ {\rm km\ s^{-1}}$, we obtain the lower limit of the maximum energy as
\begin{equation}
E^{3.0\ {\rm kpc}}_{\rm max,\ lower}\sim3.9\times10^{15}\ {\rm eV},
\end{equation}
and
\begin{equation}
E^{5.5\ {\rm kpc}}_{\rm max,\ lower}\sim6.2\times10^{15}\ {\rm eV}.
\end{equation}
Next, the upper limit of the maximum energy is estimated assuming the upstream velocity $v_{1} = 0.26 c$, which is the jet velocity around SS433 \citep{blundell2004}. We obtain the upper limit of the maximum energy as 
\begin{equation}
E^{3.0\ {\rm kpc}}_{\rm max,\ upper}\sim2.5\times10^{17}\ {\rm eV},
\end{equation}
and
\begin{equation}
E^{5.5\ {\rm kpc}}_{\rm max,\ upper}\sim4.0\times10^{17}\ {\rm eV}.
\end{equation}

Our estimation implies that the SS433 jet can accelerate cosmic-ray particles above the energy of 10$^{15.5}$ eV, although we need to investigate the appropriate value of $\eta_{B}$ by additional observations. This result is consistent with recent theoretical and numerical studies \cite{cooper2020}. Future high-sensitivity gamma-ray observation by the CTA might detect emission induced by the interaction between cosmic-ray protons and the surrounding ISM protons.

\subsection{Total Energy of Accelerated Protons}
\label{sec:tot energy of protons}
Recently, GeV and TeV gamma-rays have been detected around W50 \citep{bordas2015,abeysekara2018,sun2019,li2020,fang2020}. As mentioned in Section \ref{sec:intro}, although \citet{sun2019,li2020,fang2020} use almost the same {\it Fermi}-LAT dataset, they suggest different positions and processes for the GeV gamma-ray excess. Wherever gamma-ray emission is induced, cosmic-ray protons should interact with the surrounding ISM protons in the hadronic scenario. Moreover, the HI cavity mentioned in Section \ref{sec:HI} is most likely the origin of the gamma-ray emission. In this section, we estimate the total energy of cosmic-ray protons accelerated by the SS433/W50 system in the hadronic scenario without strictly considering the accelerating position, i.e., whether it is the spherical shell or the SS433 jet.

We define the properties of the HI cavity interacting with the relativistic protons. In some SNRs, the HI cavities have thicknesses on the scale of a few pc \citep{fukui2012,fukui2017}. Considering the proton penetration rate, it is reasonable to assume that the HI gas surrounding W50 has a thickness of 5 pc. We estimate the neutral hydrogen density of the HI gas. In the optically thin case, the proton column density can be expressed as
\begin{equation}
	N_{p}=1.823 \times 10^{18} \times W({\rm HI}) ({\rm cm^{-2}}),
	\label{eq:thin-nphi}
\end{equation}
where W(HI) is the integrated HI intensity in K km s$^{-1}$ \citep{dickey1990}. Using the HI data from 33.77 km s$^{-1}$ to 55.85 km s$^{-1}$, the velocity range of the HI cavity in the position-velocity diagram (Figure \ref{fig:figure4}), we estimate the density of the surrounding HI gas, $n^{\rm thin}_{\rm H}$, at 26 cm $ ^{-3}$ and 24 cm $ ^{-3}$ assuming distance $D$ = 3.0 and 5.5 kpc, respectively. We also consider the optically thick case. \citet{fukui2015} concluded that 85 \% of atomic hydrogen in the local interstellar atomic gas is optically thick. A way to derive the column density of optically thick atomic hydrogen has been recently constructed (see Figure 9 and Section 4.4 of \cite{fukui2017}). We derived the column density distribution from the relation between the column density and the integrated intensity of the optically thick atomic hydrogen gas. Using the HI data from 33.77 km s$^{-1}$ to 55.85 km s$^{-1}$, we estimate the density of the HI gas, $n^{\rm  thick}_{\rm H}$, as 59 cm $ ^{-3}$ and 55 cm $ ^{-3}$ for $D$ = 3.0 kpc and 5.5 kpc, respectively.

Finally, we derive the total energy of the accelerated protons. \citet{sun2019} estimated the GeV gamma-ray luminosity as
\begin{equation}
	L_{\gamma}^{pp} (0.5-10 {\rm GeV})\sim 6.48\times 10^{34} \left(\frac{D}{5.5\ {\rm kpc}}\right)^{2} {\rm erg/s}.
	\label{eq:gamma-ray}
\end{equation}
In the hadronic scenario, the gamma-ray luminosity can be written as \textit{L}$_{\gamma}$$^{pp}$$\sim$\textit{W}$_{p}$/\textit{t}$_{pp}$, where \textit{W}$_{p}$ is the total proton energy. The quantity \textit{t}$_{pp}$ is the proton-proton interaction timescale and is written as
\begin{equation}
	t_{pp}=(n_{\rm H} \sigma_{pp} \kappa )^{-1}\sim 1.9\times 10^{15} \left(\frac{n_{\rm H}}{1\ {\rm cm}^{-3}}\right)^{-1}{\rm s},
	\label{eq:pptimescale}
\end{equation}
where $\sigma_{pp}$ is the total cross section, and $\kappa$ is the inelasticity. We use the typical values $\sigma_{pp}\sim 40{\rm mb}\ =\ 4.0\times10^{-23}\ {\rm cm^2}$ and $\kappa \sim 0.45 $\citep{gabici2013}.
Therefore, $W^{\rm thin}_{p}$ can be estimated as 1.4 $\times$ 10$^{48}$ erg and 5.2 $\times$ 10$^{48}$ erg for $D$ = 3.0 kpc and 5.5 kpc, respectively. We can also estimate $W^{\rm thick}_{p}$ as 6.2 $\times$ 10$^{47}$ erg and 2.2 $\times$ 10$^{48}$ erg for $D$ = 3.0 kpc and 5.5 kpc, respectively. This energy is lower than for typical middle-aged SNRs such as W44 by one to two orders \citep{yoshiike2013}. However, the surface brightness of W50 at 1.0 GHz is higher than that of similar-scale SNRs \citep{ohmura2021}. This might be due to the difference in the scale size of the accelerating cross-section of SNRs and W50. This result implies that the gamma-ray emission originates in the SS433 jet, not the spherical shell of W50.


\section{Conclusions}
In this paper, we provided a new continuum image of W50 in the L-band. We compared the result with the observations in 1984 and 1996 and found that W50 did not experience any shape change for 33 years. There are many HI clouds around W50 within a broad velocity range. We focused on the clouds in the velocity range from 33.77 km s$^{-1}$ to 55.85 km s$^{-1}$ because of the cavity structure, which seems to coincide well with the shape of W50. We estimated the upper limit velocity of the ETF at 0.013$c$ and 0.023$c$ assuming distances of 3.0 and 5.5 kpc. We considered two energy-conservation jet models, with and without the initial spherical shell, and the observationally estimated upper limit velocities satisfy the presumed velocity based on these jet models. Furthermore, our analysis implies that the non-initial-spherical-shell scenario assuming a distance of 5.5 kpc better explains the formation of W50, although we cannot still reject the initial-spherical-shell scenario. Using the estimated upper limit velocity, we predicted the maximum energy of cosmic-ray particles accelerated by the jet terminal above the energy of 10$^{15.5}$ eV. Finally, we estimated the total energy of the accelerated cosmic-ray protons using the luminosity of the GeV gamma-ray emission. We assumed the HI cavity was interacting with W50 and estimated its density in both optically thin and thick HI gases. The estimated energies are in the range of 6.2 $\times$ 10$^{47}$ -- 5.2 $\times$ 10$^{48}$ erg depending on the density of the HI cavity and the distance of W50. These values are lower than for typical middle-aged SNRs, which implies that the GeV gamma-ray emission originates in the SS433 jet.


\section*{Acknowledgements}
We are grateful to Drs. K. Tachihara, H. Yamamoto, H. Sano, R. A. Perley, T. Tsutsumi, A. J. Mioduszewski, and V. Dhawan for useful discussion. This work was supported by JSPS KAKENHI Grant Numbers HS: 20J13339, TO: 20J12591, and MM: 19K03916. Data analysis was partly carried out on the Multi-wavelength Data Analysis System operated by the Astronomy Data Center (ADC) at the National Astronomical Observatory of Japan. This work was supported by the Japan Foundation for Promotion of Astronomy. The National Radio Astronomy Observatory is a facility of the National Science Foundation (NSF) operated under a cooperative agreement by Associated Universities, Inc. This publication uses data from the Galactic ALFA HI (GALFA HI) survey conducted with the Arecibo L-band Feed Array (ALFA) on the Arecibo 305-m telescope. The Arecibo Observatory is operated by SRI International under a cooperative agreement with the National Science Foundation (AST-1100968), and in alliance with Ana G. M\'{e}ndez-Universidad Metropolitana and the Universities Space Research Association. The GALFA HI surveys were funded by the NSF through grants to Columbia University, the University of Wisconsin, and the University of California. SAOImageDS9 development was made possible by funding from the Chandra X-ray Science Center (CXC), the High Energy Astrophysics Science Archive Center (HEASARC), and the JWST Mission office at the Space Telescope Science Institute \citep{joye2003}. This research used Astropy,\footnote{http://www.astropy.org} a community-developed core Python package for Astronomy \citep{astropy2013, astropy2018}. Mark Kurban from Edanz Group (https://en-author-services.edanz.com/ac) edited a draft of this manuscript.
\appendix
\begin{figure*}[t]
\begin{center}
	\includegraphics[width=16cm,bb=0 0 870 500]{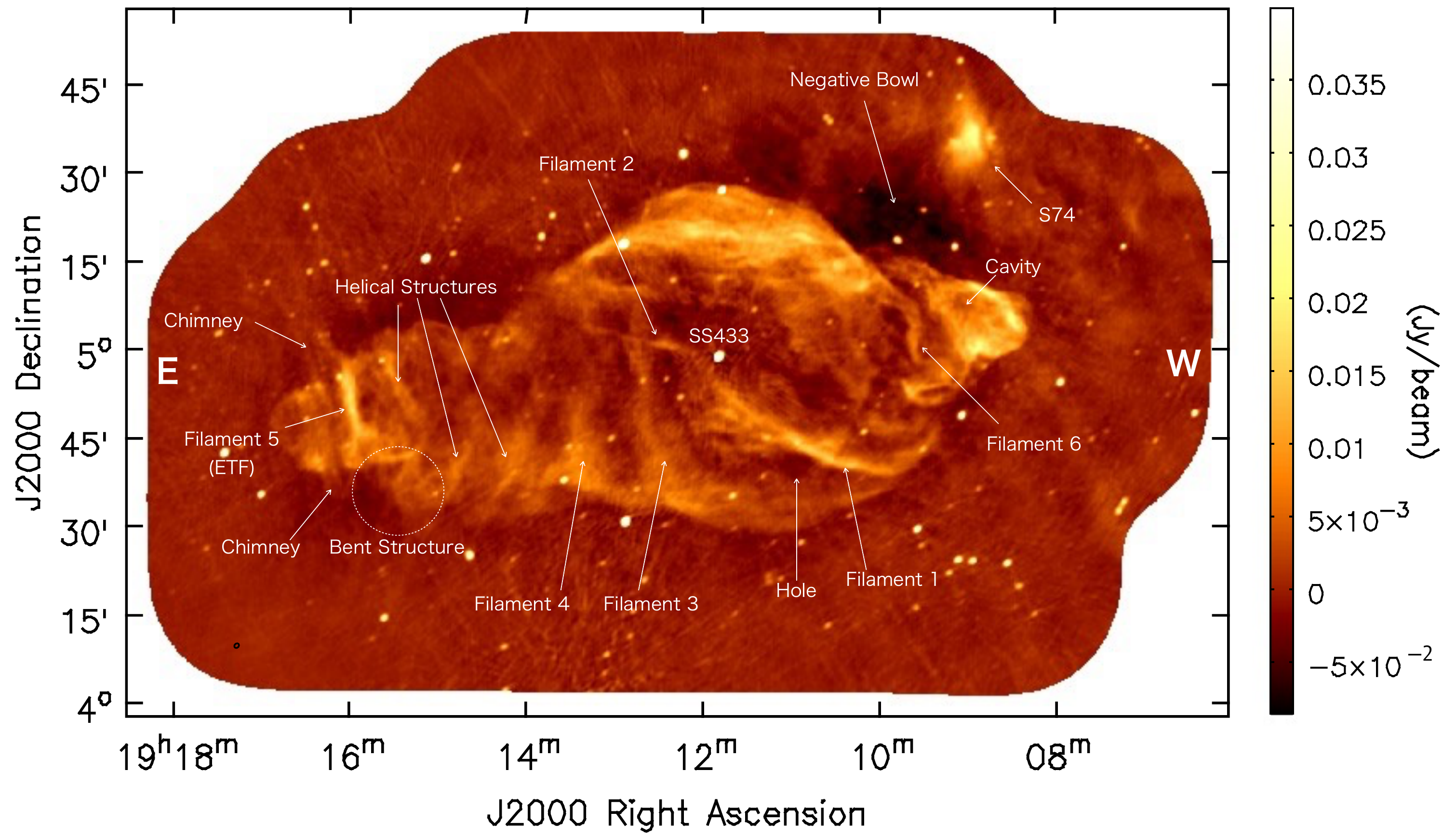}
   \caption{Total intensity map of SS433/W50. The frequency coverage is from 1.340 to 1.902 GHz. Note that the frequencies between 1.507 and 1.726 GHz are excluded owing to radio frequency interference. This image consists of 58-pointings. The ellipse at the bottom-left corner shows the beam size of 53$^{\prime\prime}$${\times}$ 43$^{\prime\prime}$. The rms noise level is 0.49 mJy beam$^{-1}$. Beside SS433/W50, there are many sources in this field of view, including bright extragalactic sources and the HII region S74 in the northwest. ``E" and ``W" indicate the east and west directions, respectively.}
    \label{fig:figure1}
    \end{center}
\end{figure*}

\section{Summary of the Radio Continuum Image in 2017}
Figure \ref{fig:figure1} shows a total intensity image of W50 observed in 2017. For clarity, we name characteristic structures and indicate them with arrows. The frequency coverage is from 1.340 to 1.902 GHz, except for frequencies between 1.507 and 1.726 GHz. Although there are artifacts around W50, they have less influence on image quality. There is also a negative bowl northwest of W50 due to a lack of the shortest {\it uv}-spacings.

We succeeded in resolving many characteristic structures of W50. The bright filament near SS433 is resolved and curves southwest (Filament 1). Filament 1 seems to be branched near SS433. Below Filament 1, there is a dark hole identified by the Low Frequency Array (LOFAR) observation at 150 MHz \citep{broderick2018}. Another faint filament is northeast of SS433 (Filament 2). It spatially coincides with the predicted rim of the helical jet, which has an opening angle of about 20$^{\circ}$\citep{margon1984}. The northern region of the central shell is especially bright and extends in the east-west direction. Some filaments seem to overlap in this region. The bright region has been suggested as resulting from interaction between the spherical shell and surrounding ISM. Because the southern shell is far from the Galactic disk, it is dimmer than the northern shell. Finally, two diffuse filaments extend from the southeast rim of the shell towards the north direction (Filaments 3 and 4). Filament 3 (the longer one) seems to consist of two overlapping filaments.

The eastern wing has a uniquely undulated shape and has several characteristic structures. The most noticeable one is a bright filament extending along the north-south direction at the eastern edge (Filament 5). We call Filament 5 the eastern terminal filament (ETF). The bright part extends to the north; however, the southern region becomes dimmer and branches off. Previous studies have resolved a faint structure called a ``chimney'' extending to the northeast region of the ETF \citep{dubner1998,farnes2017,broderick2018}. Although it is very faint, we can find the chimney in a region about 7$^{\prime}$ in size. The difference with  previous observations \citep{farnes2017,broderick2018} is that the southern chimney is clearly identified. There is a bent structure on the western side of the southern chimney, and the diameter of the eastern wing seems to change at this point. We also find unique helical structures that coil around the eastern wing. While the helical structures are similar to Filaments 3 and 4, they differ in the curve direction. 

The western wing is shorter than the eastern one because the western side of W50 is near the Galactic disk, and the jet interacts with a denser medium. There is a faint filament perpendicular to the jet axis (Filament 6) at the base of the wing. The boundary area of the western wing seems very bright; however, the inner region is a cavity. A molecular cloud is located at the cavity, and \citet{liu2020} has suggested that the molecular cloud detected by \citet{yamamoto2008} is in the western wing.

The background emission of the western area has an uneven structure owing to the Galactic disk. The HII region S74 can be seen clearly in the northwest direction from W50. There are many point sources in the entire image, which are believed to be extragalactic sources except for SS433.


\end{document}